# Hierarchies of Biocomplexity: modeling life's energetic complexity

Bradly Alicea[1]

[1]MIND Lab, Michigan State University, East Lansing, MI. USA, 48824

E-mail: freejumper@yahoo.com

**Abstract.** In this paper, a model for understanding the effects of selection using systems-level computational approaches is introduced. A number of concepts and principles essential for understanding the motivation for constructing the model will be introduced first. This will be followed by a description of parameters, measurements, and graphical representations used in the model. Four possible outcomes for this model are then introduced and described. In addition, the relationship of relative fitness to selection is described. Finally, the consequences and potential lessons learned from the model are discussed.

**1. Introduction.**

In this paper, a computational approach to biological complexity at the level of a single organism is introduced. This model might be applied to a number of domains. For example, such a model might be applied to the field of synthetic biology [1]. Currently, most synthetic biology has focused on genomic function, whereas this model would allow for a better understanding of physiological regulation. Another potential application domain is in the area of programmable artificial muscle [2], particularly in the implementation of actuators with adaptive contraction. Thirdly, this model may serve as a general model for applying computational methods to medical problems such as the complex physiology of disease states.

The theory of evolution depends on natural selection and neutral processes, which in turn depend upon interactions at multiple scales of analysis. As a secondary aim, I will take the position that the effects of environmental selection on organismal[1] physiology can be abstracted to a series of units that replicate at a specific level and are interconnected with units at other levels. This relates to outstanding controversies in multilevel selection, as a classic problem is approached using systems-level and computational approaches.

*1.1 Introduction to problem.*

Variation is the raw material of evolution. In biological systems, this variation is ubiquitous, existing at every scale of analysis. It can also be argued that the self-organization of biocomplexity is the product of two phenomena: natural selection and the existence of multi-scalar phenomena. To better understand how these phenomena are central to determining the dynamics of self-organization, the levels of selection problem and the concept of hierarchical energetic organization will be brought to bear on the problem.

*1.2 Levels of selection.*

While the levels of selection problem is a controversial topic, the idea of selection acting at multiple levels of biological organization can be informative for understanding the structure of biocomplexity. Selection can act at multiple levels, from molecular to phenotypic to cultural [3]. When selection acting at one scale is observed at another scale, its effects may appear to be diffuse. The magnitude of this effect is dependent on the units of selection at a certain level. Dawkins [4] conceives of these units as replicators. These replicator units will become important in understanding the structure of each level and interactions between levels.

---

[1] Organismal can be defined as having to do with the properties of the organism.

*1.3 Introduction to scale.*

Intuitively, a difference in scale refers to an order of magnitude. The concept of scale in biocomplexity has been explored using a number of different examples and perspectives [5, 6]. Most recently, this approach has been applied to understanding physiological phenomena [7]. In the abstraction presented here, a single level can be better understood as a set of scalar elements (see Figure 1) with two properties: 1) the elements of a single scalar set are embedded in a hierarchy of scalar sets, and 2) each scalar set contains multiple replicators.

Thinking of the levels of selection problem in this way may resolve a central contention in modern conceptualizations of multi-level selection: how does selection at a particular scale affect features at other scales? In other words, how does group selection affect the relative fitness of an organism, or how do changes in gene expression and phenotype interact? These questions are especially important when we consider that elements at a lower scale (e.g cell populations) interact with each other to form emergent properties at higher scales (e.g. organs).

*1.4 Biocomplexity reconsidered.*

In the sense of making the connection between genomes and ecology, biocomplexity is inherently multi-scalar (see Figure 1). An organism, from its genes to its ecological surroundings, is organized in a hierarchical fashion [8]. More importantly, these hierarchical levels are interconnected in the fashion of a bi-directional complex adaptive network. Therefore, living systems should be expected to exhibit phenomena such as emergence or self-organized criticality [9]. Furthermore, natural selection should drive these phenomena via activational hormone spikes during development, population structure in evolution, and paracrine signaling to trigger transitive gene expression across life-history [10].

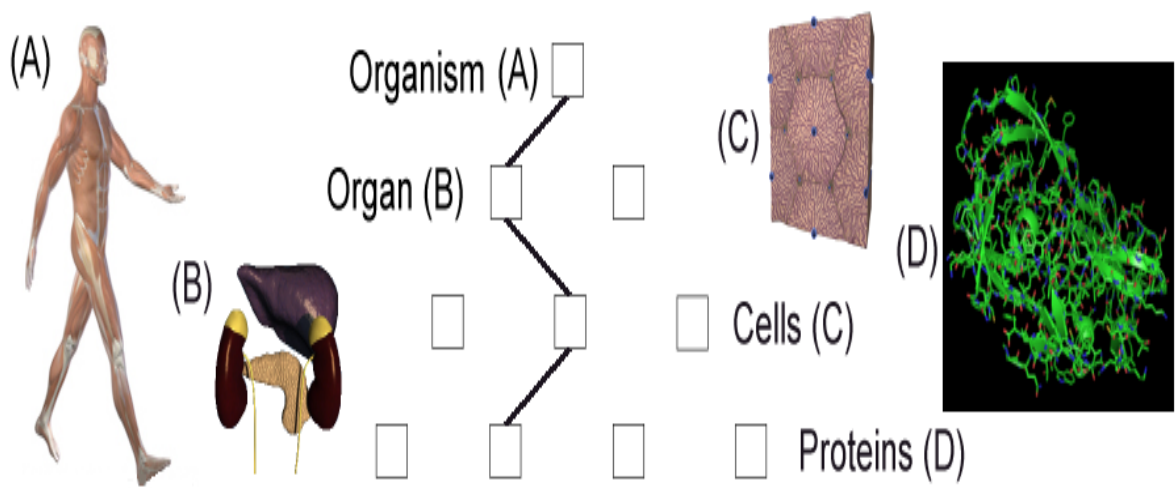

**Figure 1.** Diagram of the multiscalar nature of an organism. A) lateral perspective of a human organism and its musculoskeletal system, B) liver/kidney/pancreatic organ system, C) population of liver cells, D) model of HIV-1 protease[2].

In this model, biocomplexity is not only defined as a formal hierarchy, but also as a complex, interconnected system with energetic structure. The connectivity between different hierarchical scales, which is important in the unfolding of emergent phenomena, is very much controlled and enabled by the flow of energy. All living organisms are open systems, meaning that there are continuous flows of energy available from outside the organism. The amount of free energy at any given time is subject to the first law of thermodynamics[3] and determines the thermodynamic potential of the system [11]. This pool of energy is subject to constant fluctuations given processes at multiple hierarchical scales. Furthermore, the relative amount of free energy over time can drive competition between replicators. This multi-scalar, energetic aspect of biocomplexity might very well be the "hidden" aspect of natural selection.

---

[2] NOTE: pathway has been drawn through the hierarchy to describe the location of the diagrams in the system.

*1.5 Critical effects.*

Selection may act upon any level of the hierarchy, but the actual perturbation applied (e.g. the intensity of selection) may vary in terms of scope. In the case of UV damage, selection applied at the cellular level may affect a single cell [10]. In the case of certain cancers [13], entire cell population and organs are affected. Yet other phenomena, such as the "lock and key" relationship between proteins and their receptors, bridge scales and differentiate between replicator units at single scales [14]. The effects of this selection pressure will vary and be translated to other hierarchical levels accordingly. In addition, there can be synergistic effects between scales. For example, lactose tolerance evolves as a combination of cultural and environmental pressures and changes in gene expression during life-history. There are three variables that determine the critical effects of either sexual or environmental selection on the hierarchical structure of life: scope, connectivity, and transformity[4].

**2. Methods/Concepts.**

Concepts from systems ecology, complexity theory, and systems biology can be used to better understand relationships between the different scales of analysis that exist in an organismal physiology. These concepts are the conservation of energy, trophic structure, connectivity between hierarchical scales, and the relationship between transformity and selection.

*2.1 Conservation of energy.*

In the proposed model, energy is conserved at a rate of $10^{-n}$, where $n$ is the level of transformity. Transformity can be defined as the amount of energy conserved from one hierarchical level to another [15]. A single level of transformity is equivalent to a single trophic relationship in an ecological system.

*2.2 Trophic structure.*

As in the food web of an ecosystem, each successively higher level of the physiological hierarchy is represented by an order of magnitude fewer constituent components. For representational purposes, different hierarchical levels forms a single functional system by means of a directed hierarchical graph with feedback, but this model permits non-nested sets, which can be defined as hierarchical connections that exhibit connectivity exceeding that of a strict hierarchy.

*2.3 Scope.*

The scope of multilevel selection has to do with how wide ranging is the selection pressure at a particular level. Scope can be defined mathematically as

$$S = U_{hs} / U_{hns} \qquad [1]$$

where $S$ is the scope, $U_{hs}$ is the number of units at a specific scale under selection, and $U_{hns}$ is the number of units at $10^{-n}$ not under selection. When considering the dynamics of selective scope, scope (S) values at or near 1 result in a vast scope, while scope (S) values at or near 0 equal a paucity of scope.

---

[3] In this context, the first law of thermodynamics refers to the following: during any transformative process, energy is neither created nor destroyed. All energy takes the form of either free energy (available to do work), or bond energy (the product of entropy and thus not available to do work).
[4] Transformity can be defined as the scaling factor at which energy is transferred between hierarchical scales.

*2.4 Connectivity.*

Whereas scope can be defined as the extent of selection's effects on replicators at each level, connectivity can be defined in relation to different levels of hierarchical scale. Connectivity provides an assessment of how many replicators at the next highest hierarchical scale ($10^{-n+1}$) are connected to a single replicator at the hierarchical scale of observation ($10^{-n}$). Connectivity can be defined mathematically as

$$C = 1 - \Sigma (U_n / C_u) \qquad [2]$$

where C is the connectivity, $U_n$ is a single unit at the hierarchical scale of interest, and $C_u$ is the number of connections to replicators at other hierarchical levels. Higher values for the connectivity measure translate into greater connectivity between scales and potentially more widespread effects of selection at lower scales.

*2.5 Transformity and selection.*

In this model, the outcome of selection has a quantifiable effect based on energetic principles. The amount of transformity for selection imposed at a specific hierarchical scale involves a certain amount embodied energy, or emergy, derived from the transfer of energy between hierarchical scales. Examples of this transfer include processes such as transcription, macromolecular assembly, intercellular communication within an organ, and the energy required to maintain sociality. Embodied energy can be defined as an estimate of both the amount of Gibbs free energy[5] and bond energy (e.g. entropy) that goes into a certain process. This can be defined mathematically as

$$E_n = G_n + B_n \qquad [3]$$

where $E_n$ is the embodied energy for replicator unit *n*, $G_n$ is the Gibbs free energy for replicator unit *n*, and $B_n$ is the Bond energy for replicator unit *n*. In general, higher hierarchical levels have a greater emergy per unit than lower levels. In this model, selection can be defined as the strength of a behavioral or physical signal that puts stress on the organism or group. This produces effects that have a larger direct energetic effect when imposed at higher levels.

*2.6 Fitness as a function of selection.*

Fitness is calculated as a relative measure between different replicator unit populations at each hierarchical scale. In order to arrive at a suitable fitness measure relative to the entire hierarchy, we must first calculate two additional variables: pervasiveness (*P*) of the replicator unit in question, and a relative embodied energy ($E_r$). Pervasiveness can be calculated using the following equation

$$P = 0.5 * (S + C / C_{max}) \qquad [4]$$

where *S* is the scope from eq. 1, and *C* is the connectivity from eq. 2. $C_{max}$ is the maximum connectivity

---

[5] The mathematical definition of Gibbs free energy can be found in [11], Page 431.

value in the entire hierarchy. Relative embodied energy can be calculated by the following equation

$$E_r = (E_n - E_{min}) / (E_{max} - E_{min}) \qquad [5]$$

where $E_n$ is the embodied energy calculation from eq. 3, $E_{max}$ is the maximum embodied energy in the entire hierarchy, and $E_{min}$ is the minimum embodied energy in the entire hierarchy.

Relative fitness can be defined verbally as the energetic value and aggregate effect of a particular replicator unit or unit populations on the entire hierarchy relative to all other units in the hierarchy. Relative fitness can be defined mathematically as

$$F = E_r + P / 2 \qquad [6]$$

where $E_r$ and $P$ are from eqs. 4 and 5, respectively. The value of F can range between 0 and 1 so that replicator units or unit populations with a high embodied energy and large aggregate effect on the rest of the system have a value approaching 1.

## 3. Results.

The results of this method can best be understood graphically considering the system in equilibrium. Recall that there are two components to the model: units of selection (e.g. replicators), and potential connectivity between the different scales of analysis. Furthermore, each scale of analysis has a unit of energetic conservation (e.g. transformity) that roughly describes the effects of selection and normal functional processes as the complex phenotype emerges from lesser scales. The first result is to graphically represent the replicators and units of energetic conservation.

Figure 2 graphically demonstrates the hierarchical relationship involving each biological scale while ignoring connectivity. When connectivity is taken into account, four scenarios should be expected: selection imposed at a lesser scale with a relatively large effect on the phenotype, selection imposed at a greater scale with a relatively small effect on the phenotype, selection imposed at a greater scale with a relatively small effect on the phenotype, and selection imposed at a greater scale with a relatively large effect on the phenotype.

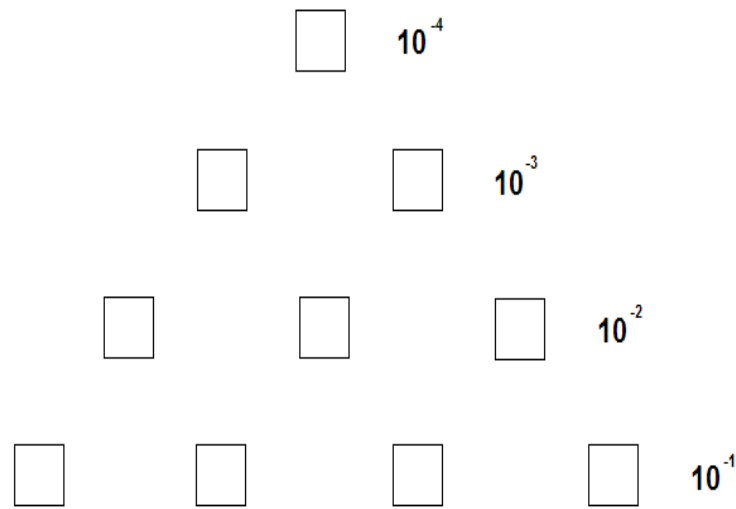

**Figure 2.** A non-interconnected example of hierarchical organization using generic units (replicators) to represent entities at each scale.

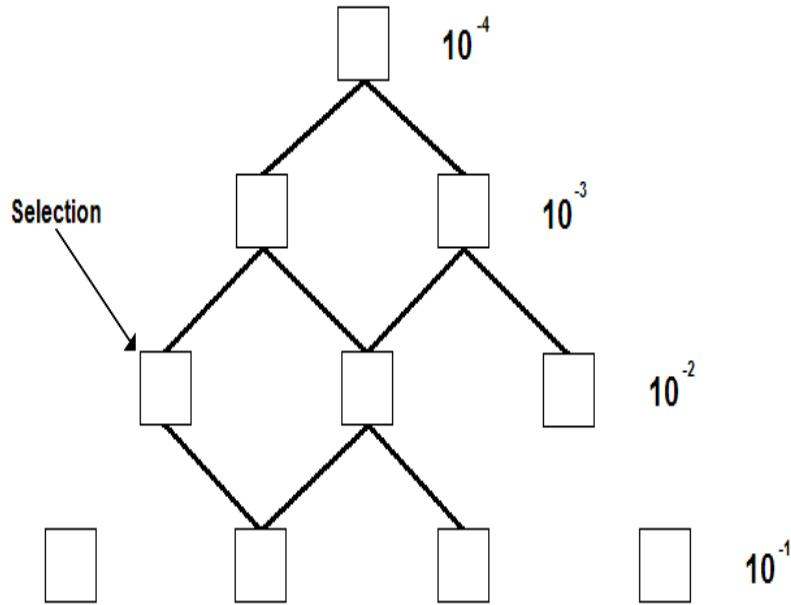

**Figure 3.** An example of selection acting on a lower hierarchical level of the organism with a relatively large effect.

*3.1 Four potential outcomes of model.*
   In this model, selection is introduced at one or more replicator units at a particular hierarchical scale. In the examples shown in Figures 3 through 6, the level and scope of selection is varied with respect to how it affects individual replicator units.

   In Figure 3 and Table 1, selection is imposed at a lower hierarchical level, while its effect is relatively large. In terms of the potential values of the parameters, the scope value should approach 1, the transformity value should approach $10^{-1}$, and the connectivity value should be relatively small. In Figure 4 and Table 1, selection is imposed at a lower hierarchical level, while its effect is relatively small. In terms of the potential values of the parameters, the scope value should approach 0, the transformity value should approach $10^{-1}$, and the connectivity value should be relatively large. In Figure 5 and Table 1, selection is now imposed at a higher hierarchical level, while its effect is relatively small. In terms of the potential values of the parameters, the scope value should approach 0, the transformity value should approach $10^{-max}$, and the connectivity value should be relatively large. In Figure 6 and Table 1, selection is imposed at a higher hierarchical level, while its effect is now relatively large. In terms of the potential values of the parameters, the scope value should approach 1, the transformity value should approach $10^{-max}$, and the connectivity value should be relatively small.

*3.2 Feedback across scales of replicators.*
In this model, the flow of energy always proceeds from lower hierarchical scales to higher ones. For example, 37 units of energy at one scale is equivalent to 3.7 units at the next highest scale, and .37 two scales higher. The energetic scaling is roughly .10, which is consistent with ecological systems [15]. When selection is imposed at a particular scale, it directly affects the units under selection at that scale and potentially all units at lesser hierarchical scales. However, selection also has indirect effects on connectivity and the biocomplexity hierarchy as a whole.

   In this model, selection affects individual units in a subtractive manner, taking away from their embodied energy from greater to lesser hierarchical scales in a cascading fashion. Connectivity is then affected by this loss of embodied energy. When a given unit drops below a specific threshold for embodied energy, connectivity between it and other units can cease to exist depending on the connectivity and embodied energy of the other unit. This is demonstrated in various forms for Figures 2 through 6.

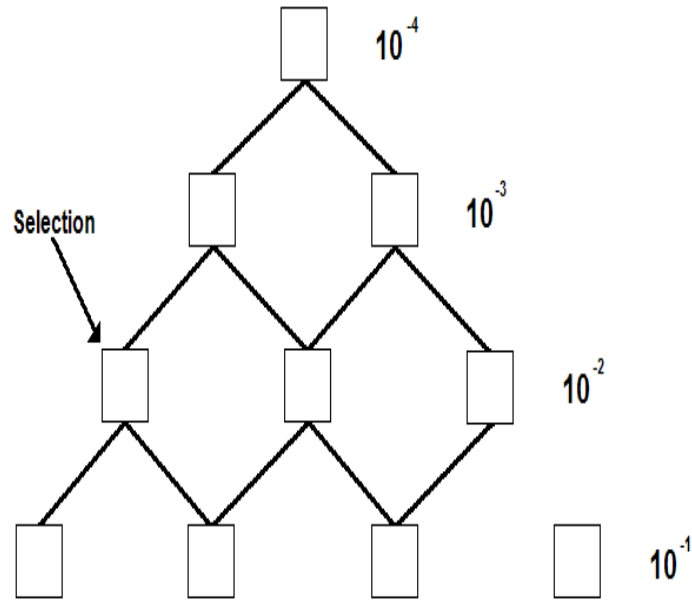

**Figure 4.** An example of selection acting on a lower hierarchical level of the organism with a relatively small effect.

**Table 1.** Values for scope (S), transformity (T), and connectivity (C) in each condition.

| Selection Imposed, Magnitude of Effect | S | T | C |
|---|---|---|---|
| **Lower, Large** | ~ 1 | ~ $10^{-1}$ | small |
| **Lower, Small** | ~ 0 | ~ $10^{-1}$ | large |
| **Higher, Small** | ~ 0 | ~ $10^{-max}$ | large |
| **Higher, Large** | ~ 1 | ~ $10^{-max}$ | small |

*3.3 Energy transfer between scales.*

When selection is applied to a particular scale, it has a specific energetic effect on the organismal hierarchy. In a general sense, selection imposed at one unit has a feedback effect throughout the network (see Figure 7). Figure 7 demonstrates how selection is quantified as a specific energetic value. Frame A shows the energetic value of one unit at the higher scale and two units at the lesser scale. Frame B shows the contribution of energy from the lesser scales to the higher scale and the selection pressure on the higher scale. In this case, selection is negative and acts as a feedback mechanism to energy flow in the system. Frame C shows how this feedback mechanism indirectly affects the lesser scale by passing on embodied energy at a factor of $10^{-1}$. These three steps are repeated over time and can be extended to the rest of the network topology

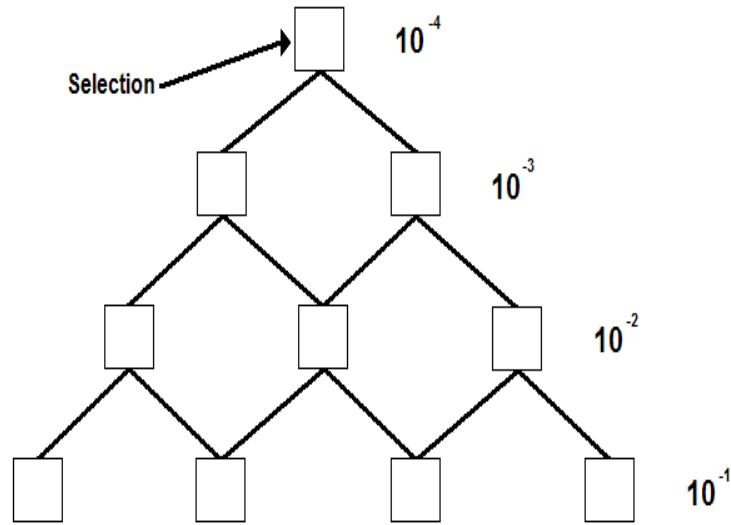

**Figure 5.** An example of selection acting on a higher hierarchical level of the organism with a relatively small effect.

*3.4 Notes on dynamic aspects on model.*

There are three factors to keep in mind that relate to the dynamic aspects of this model. These were discovered during the course of setting up and running test models. These are: competition and growth of replicator units, the self-limiting proliferation of units at lower scales, and the definition of units as discrete entities. Each of these factors will affect the overall robustness of the model.

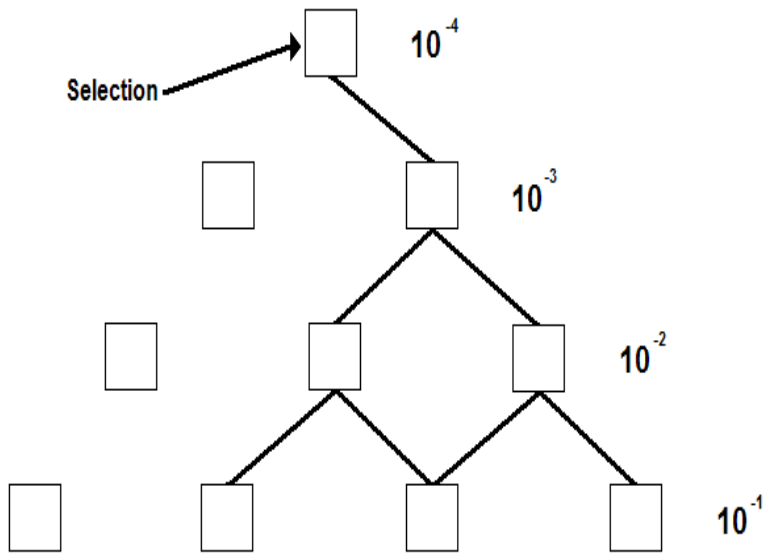

**Figure 6.** An example of selection acting on a higher hierarchical level of the organism with a relatively large effect.

*3.5 Competition at a single scale.*
   The first factor involves competition and growth of replicator units at single scales of analysis. In a purely biological context, this corresponds to different cell types or rate-limiting processes such as gene regulation. In various scenarios, we should expect that some replicator unit groups will overtake others in terms of number, and that these groups will out-compete each other for energy flowing through the system.

*3.6 Self-limiting proliferation of units.*
   The second factor involves the self-limiting proliferation of units at lower scales. In Figures 2 through 6, we have so far considered a hierarchy where there are a progressively greater number of units the farther down the hierarchy one travels. This type of structure works well for modeling the progression from organism to proteins. However, at the scale of genes, this becomes problematic.

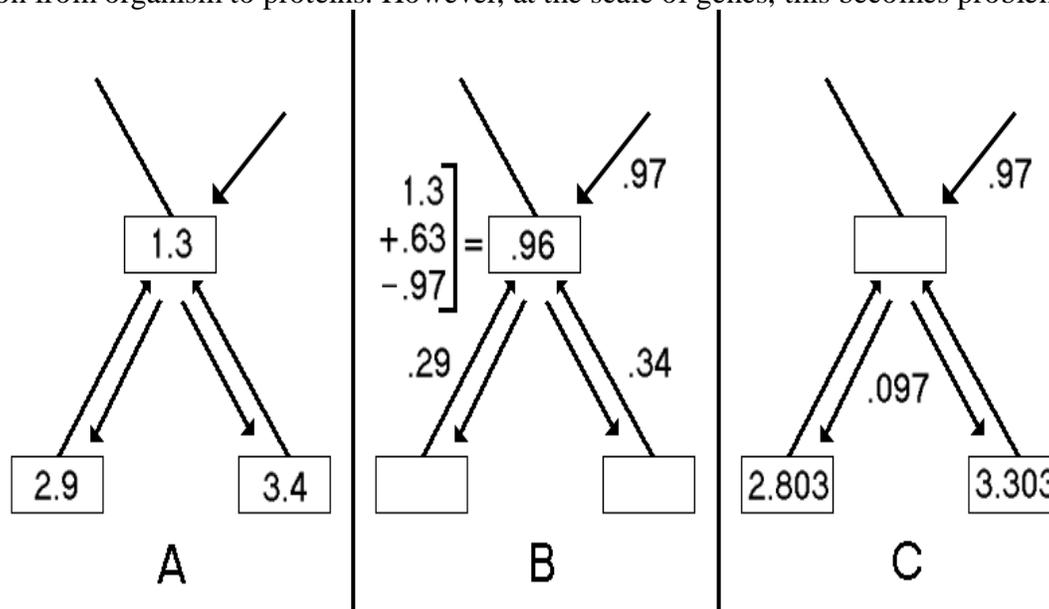

**Figure 7.** Basic model of energetic transfer (e.g. transformity) between scales.

   If we consider the relationship between genes, gene products, and cells, it becomes clear why this must be accounted for: while there are only about 20 to 25,000 protein-coding genes in the human genome [16], the expression of these genes is combinatorial. Furthermore, the number of distinct transcripts may or may not necessarily larger than the number of cells in the organism.

   This caveat regarding the proper number of units per scale brings us to the third factor. It is important to consider how the individual units are defined. This is true in terms of both which biological phenomena are selected at each scale and how hierarchical each phenomenon exists in an actual biological context.

   Finally, replicator unit population at any scale should be expected to be limited by energy flux dynamics and homeostatic balance. This can be independent of competition with other replicator unit populations, and is related to the pervasiveness ($P$) of the unit population. In cases where the value for $P$ approaches 0, the population should be expected to be severely limited by energy flux. Conversely, $P$ values close to 1 should not be constrained by energy flux, and also exhibit a high relative fitness.

*3.7 Relationship between selection and fitness.*
   Figures 8 and 9 are schematics using pseudodata to show the relationship between selection and the fitness of the entire hierarchical structure. Figure 8 shows a fitness surface landscape in relation to a given magnitude of selection and hierarchical scale. In application, for every selection and hierarchical scale combination there should be a corresponding relative fitness. This type of fitness landscape should also help to predict results that are more or less likely given certain conditions. Figure 9 shows the bivariate relationship of selection against relative fitness. The four curves (A through D) show the relationship between selection and relative fitness for different replicator unit populations. When curves A and D are compared, it is shown that the fitness of A decays to 0 given a much lower level of selection applied to the system. This may be due to differences in the energetic baseline, scope of selection, and connectivity between replicator unit populations.

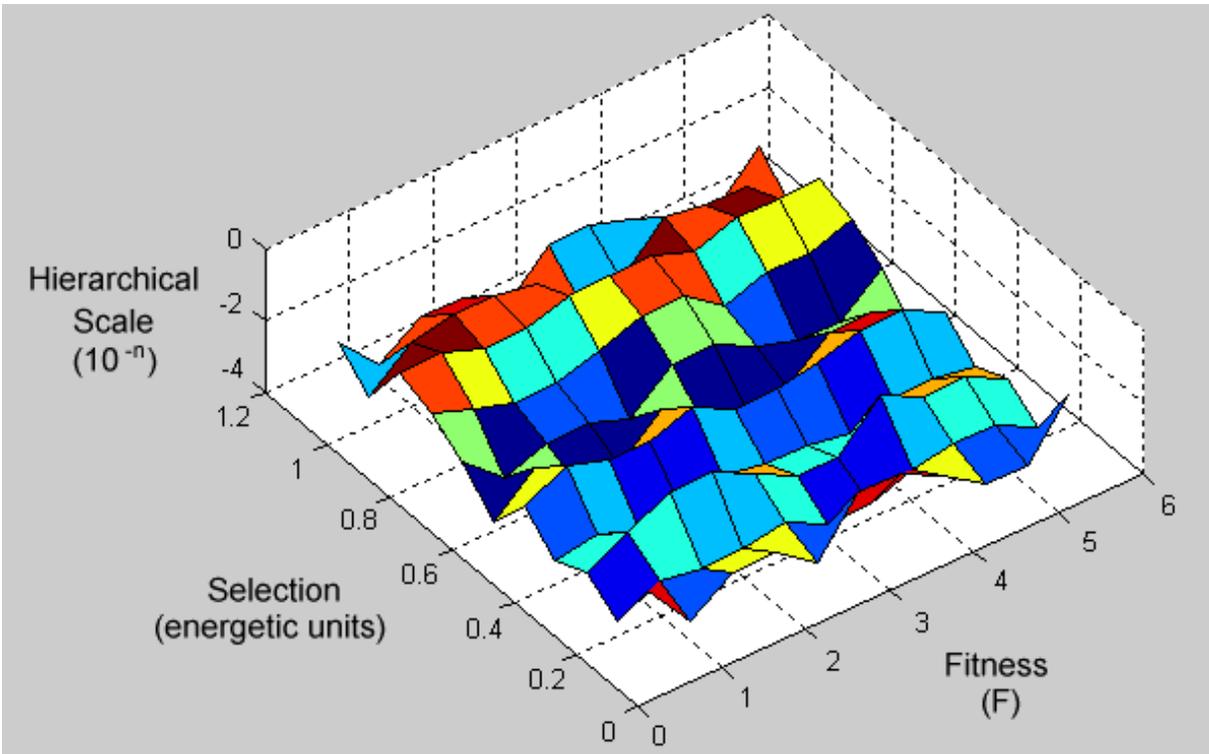

**Figure 8.** Relationship between selection, relative fitness coefficient, and hierarchical scale.

## 4. Discussion and Conclusions.

One of the lessons conveyed here is that living systems are self-organized entities. Therefore, the structure of relationships between components at different scales should be expected to vary across development and other phases of life-history. In addition, life-historical and other processes that unfold as the organism or specific biological systems are active will "rewire" the hierarchical network, making the effects of selection truly a dynamic process.

*4.1 Take-home messages.*

Selection can have amplified or dampened effects on distant hierarchical scales. Related to this, there are two phenomena which also affect the structure of biocomplexity hierarchies: competition between replicator units, and epistasis[6] or epistatic-like interactions [17] at multiple scales. In many cases, the aggregate effects of selection are shaped by these processes.

---

[6] Epistasis can be defined as the quantifiable interactions between genes. In general, epistasis can be antagonistic, additive, or synergistic [17], which results in different patterns of interaction at higher hierarchical scales.

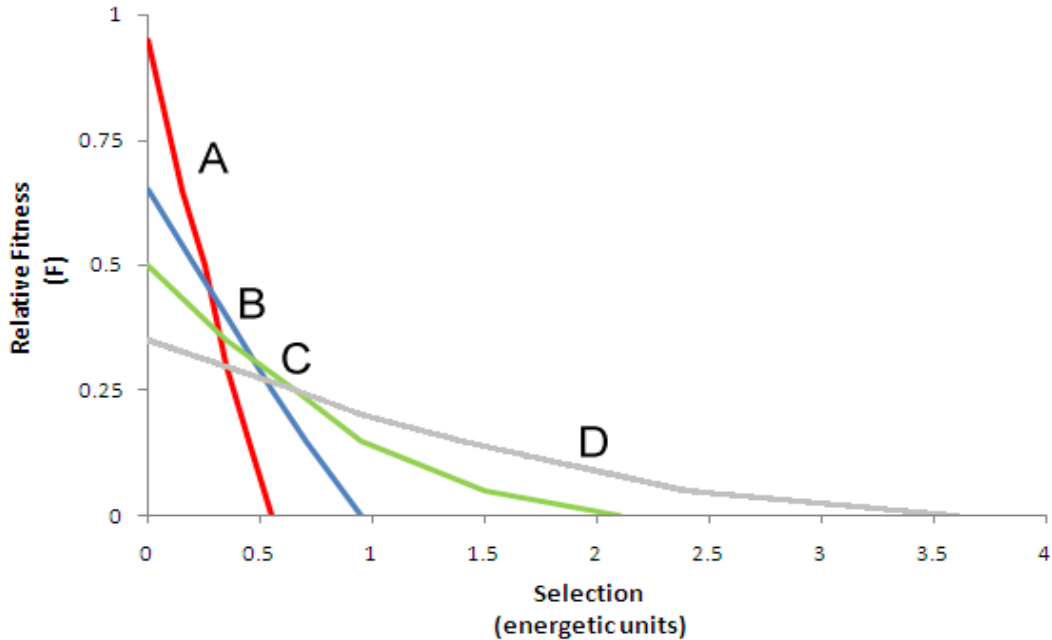

**Figure 9.** Curves representing the relationship between relative fitness coefficient and selection. Curves A through D graphically demonstrate the relationship between selection and relative fitness for different replicator unit populations.

*4.2 Competition between replicator units.*

Upon reviewing this model, one might think that all replicator units at a single hierarchical scale are homogeneous. This is not always true. In fact, there may be competition between different types of replicator units that approximate biological processes such as the spread of cancer, cell differentiation, or competing organismal populations. Typically, these competing populations are characterized by a separable connectivity meaning that connection from and to these units are to be clustered by function.

*4.3 Epistasis at multiple scales.*

One consequence of organizing each scale as a series of parallel replicators is that epistasis [18, 19], or the gene-gene interactions that define complex output at the genotypic level, also plays a role at higher scales of analysis. In this case, replicators with a high degree of connectivity (C) at one scale will overlap in terms of their effect on units at higher scales. One concrete example would be individual cells can have a high degree of influence on other cells in their cell population, and even on the organ as a whole. Most synonymous with epistasis is the influence of individual cells on a higher level structure (in this case the cell populations) can be either additive or multiplicative. This can act to amplify the effects of selection, especially if the scope of selection is large.